\documentclass[a4paper,11pt]{article}
\usepackage{jinstpub} 
\usepackage{lineno}
\usepackage{subfig}
\usepackage{comment}

\usepackage{longtable}
\usepackage{lipsum} 

\notoctrue

\DeclareOldFontCommand{\rm}{\normalfont\rmfamily}{\mathrm}
\DeclareOldFontCommand{\sf}{\normalfont\sffamily}{\mathsf}
\DeclareOldFontCommand{\tt}{\normalfont\ttfamily}{\mathtt}
\DeclareOldFontCommand{\bf}{\normalfont\bfseries}{\mathbf}

\title{Discovering the underlying analytic structure within Standard Model 
constants using artificial intelligence}

\author[a]{S. V. ~Chekanov} 
\author[b]{and H.~Kjellerstrand}

\affiliation[a]{HEP Division, Argonne National Laboratory, USA}
\affiliation[b]{Hakank.org, Sodra Forstadsgatan 40b, 21143 Malmo, Sweden}

\emailAdd{chekanov@anl.gov, hkjellerstrand@acm.org}

\abstract{
This paper presents a method for uncovering hidden analytic relationships among the fundamental parameters of the Standard Model (SM), a foundational theory in physics that describes the fundamental particles and their interactions, using symbolic regression and genetic programming.
Using this approach, we identify the simplest analytic relationships connecting pairs of these constants and report several notable expressions 
obtained with relative precision better than 1\%. 
These results may serve as valuable inputs for model builders and artificial intelligence methods aimed at uncovering hidden patterns among the SM constants, or potentially used as building blocks for a deeper underlying law that connects all parameters of the SM through a small set of fundamental constants.
}


\note{Preprint: HEP-ANL-197373, June 26, 2025}

\begin{document}

\maketitle
\flushbottom

\clearpage

\section{Introduction}
\label{sec:intro}

The Standard Model (SM) is the theory in physics that describes the fundamental particles and their interactions. It remains highly successful in predicting a wide range of experimental results. However, it is still considered a model rather than a complete theory, as it relies on approximately 20 fundamental parameters that must be determined experimentally \cite{ParticleDataGroup:2024cfk}. 
For example, the masses of the most fundamental particles---such as the six quarks (denoted by the letters $u$, $d$, $s$, $c$, $b$, and $t$), the vector bosons ($Z$ and $W$), the Higgs boson ($H$), and the leptons (electrons $e$, muons $\mu$, and tau leptons $\tau$)---are treated as free parameters that must be determined experimentally. This reliance on such parameters limits the predictive power of the SM. One of the most important open questions in the SM is why these constants have the specific values we observe, and whether a deeper theoretical framework could explain them.

The Large Hadron Collider (LHC) has enabled numerous stringent tests of the SM, yet no evidence of new physics has been observed. Most searches for new physics at the LHC rely heavily on the SM’s predictive power. However, the SM contains numerous adjustable parameters that must be determined experimentally, which limits the interpretability of the measured cross sections. 
Moreover, the SM offers limited predictive power for many categories of events produced in $pp$ collisions, leaving significant regions of event topology still largely unexplored \cite{universe10110414}.

Attempts to identify relationships among the parameters of the SM have a long \mbox{history \cite{WBook, Nielsen:1994ab, Froggatt__2003}.} However, no convincing evidence has been found to support the existence of such relationships. As a result, these constants are generally regarded as free parameters of the SM, with no apparent underlying connections.

When there is no available theory that can describe apparently unconnected constants, the first step is to identify the simplest analytic relationships  between their values, before constructing a theory based on specific physics principles. 
Finding analytic relationships through brute-force searches by noticing coincidences in measured values is common for science---even before the development of fundamental theories that could explain connections between seemingly unrelated quantities. This empirical approach proved to be useful and sometimes led to discoveries of new laws that revealed the underlying reasons for such coincidences. 
For example, in 1885, Johann Balmer derived an empirical formula that accurately described the wavelengths of the visible spectral lines of hydrogen. At the time, it was regarded solely as a mathematical relation, lacking a corresponding physical interpretation.   But this mathematical regularity was later explained by quantum mechanics through Bohr’s model of the atom in 1913. 

A similar strategy may be valuable for studying the many free constants of the SM. Compared to the earlier example, the main challenges lie in the large number of parameters, the presence of physical units for some of them, and the wide variation in measurement precision, since some constants are known with extremely high accuracy, while others are not. Today, however, the search for numerical regularities in many seemingly unrelated parameters is becoming increasingly feasible thanks to advances in artificial intelligence (AI) and growing computational power.

In this paper, we present an algorithm designed to uncover potential underlying structures among the fundamental constants of the SM using symbolic regression and genetic programming (GP)---a type of evolutionary algorithm inspired by natural selection and rooted in the broader field of AI. We apply this method to explore whether analytic relationships among SM constants may hint at an intrinsic structure or possibly point toward a unifying law that governs them.
We also discuss how our method may be adapted for application in a broad range of scientific fields.

\section{The Method}
\label{sec:method}

Our AI-based approach to addressing the problem of free parameters in the SM proceeds as follows:

\begin{enumerate}
    
\item
The first step is a traditional symbolic regression \cite{koza92}, which derives analytical expressions from SM parameter data, generating a large set of candidate relations.
This step is the most computationally demanding and produces large datasets containing analytical relationships among SM constants.
Such datasets inevitably include spurious analytic expressions that arise purely from numerical coincidences rather than genuine physical connections.

\item
The second step operates on the datasets generated in the first stage. At this stage, dimensional analysis and general expectations from the SM are applied to identify the most plausible analytical structures within a high-dimensional functional space. 
We expect that some constraints can be applied to filter out numerical noise and retain only relations that satisfy certain principles.
The obtained relations may provide hints of symmetries or dynamical models underlying the SM. However, since dimensional analysis has not yet been implemented in the GP algorithm, this step cannot currently be performed in a straightforward manner. 

\end{enumerate}

This paper describes the first step in full detail, i.e., the creation of datasets of analytic relationships between the SM parameters generated by the GP algorithm.
The purpose of this step is to provide analytical relationships among fundamental physical constants, some of which can be used as building blocks for models with a reduced number of free parameters.
These datasets can also serve as input for various AI techniques to analyze patterns and structures that may reveal hidden relationships among SM parameters. 
We will also provide an example of the second step based on some basic SM expectations. However, a full analysis involving this step should be presented in a dedicated physics analysis paper, which is beyond the scope of this article.


\section{Technical details}
\label{sec:simulation}

Genetic programming is a powerful technique in symbolic regression that evolves analytic expressions directly from data \cite{koza92}. By representing mathematical expressions as trees and applying evolutionary algorithms, GP explores a wide solution space unconstrained by predefined model structures, enabling the discovery of complex, previously hidden relationships. 

In symbolic regression, the system learns an (symbolic) expression from a set of input--output pairs, often many such pairs, and often there are many input values (representing the dependent variables), and a single output (the independent variable). The generated expressions then contain relations with these variables, functions of these variables, as well as constants. It can then be used to evaluate expressions with inputs not seen in the given data set. 

The symbolic computations were performed using a program written in the {\sc Picat} language \cite{PicatWebsite}, a logic-based, multi-paradigm programming language that combines features from logic programming, constraint solving, functional programming, planning, and imperative programming. The program primarily leverages {\sc Picat}’s logic programming (non-determinism), functional constructs, and imperative features. It is available on the second author’s website \cite{PicatSymbolicRegression}.

The specific variant of symbolic regression used here is designed for constant identification. This method is different from the traditional symbolic regression in some important aspects:
  \begin{itemize} 
  \item The data set for a single run/experiment  consists of a single input-output pair:  $(a,b)$,  where $a$ is an input value, and $b$ is a value of the expected output.
  \item The algorithm then tries to learn some relation between the constant $a$ and $b$, generating an analytic expression. See below for more on the specific genetic programming method.
  \item The found expression does not contain variable names as described above for the general symbolic regression approach. In our version of constant identification, everything in the generated expression is either a function or constants (a physical constant or an integer). 
\item The equation for the physical constants that is shown below in the paper contains the name of the physical constants, but that is handled by post-processing after the symbolic regression part is completed.
\end{itemize}

Some specific features of the program are:
\begin{itemize}
  \item The number of expressions in each population is 1000. In the first run, the expressions are generated randomly from:
  \begin{itemize}
    \item functions: +, $-$, *, /, $\sqrt{x}$,  $x^2$, $x^3$, $x^4$, $\exp(x)$, $\log(x)$, $\cos(x)$, $\sin(x)$, $\tan(x)$, $\arctan(x)$;
    \item the integers from 1 to 10. The use of only small integers instead of larger integers or random floats is for simplicity and for the interpretability of an expression;
    \item and the two or three physical constants involved, the output and input value(s).
  \end{itemize}

  \item 
  The analytic formulas are evaluated for each expression in a population, and the results that are closest to the output value are kept for the next generation. The formulas that are too far from the output value are replaced with previous expressions that are combined from other expressions using the standard genetic programming operators crossover and mutation (with some specific probability). New random formulas may also be added (from the same set as shown above).
  \item When an expression is sufficiently close to the value of output $c$---within the allowed error of the output---then this is considered a solution. All solutions found within the allotted number of runs (1000) are reported. If no solution is found within the allotted number of runs, the program continues to run until a solution is found (and then reported), or until the program times out. The timeout for each experiment, based on a given output--input configuration, was set to 60 s.

\item 
An important feature of the program is that it enforces the inclusion of all input value(s) in each expression. Any expression that does not contain the required input value(s) is removed from the population.

\end{itemize}

The results presented in this paper are based on 3249 distinct runs using inputs either with or without mass units. Each involving the testing of an output against one or two input constants, as will be discussed later. Many single runs produced multiple solutions. The results were obtained from multiple runs with different random seed values, which takes into account the inherent stochastic nature of GP. Duplicate solutions were removed during the post-processing step. All runs took, in total, about 40,000 CPU hours with the Intel E5-2650 v4 processor and 252~Gb RAM.

\subsection{Inputs for the genetic algorithm}

A critical aspect of evaluating analytic relationships between SM parameters is the selection of an appropriate input space for the algorithms. 

Working with SM parameters presents several challenges. First, most of these parameters have physical units, which restricts the applicability of the algorithms due to dimensional analysis constraints---only certain analytic structures are valid when dimensions are taken into account.

\subsection{Inputs with Units}

Table~\ref{tab:table1} lists the fundamental SM parameters commonly used in particle physics, as reported by the PDG \cite{ParticleDataGroup:2024cfk}. 
These constants, including the masses of fundamental particles expressed in physical units (e.g., MeV), are used as inputs for GP. We will discuss these constants below.

We did not include the PMNS mixing angles \cite{ParticleDataGroup:2024cfk} in the GP algorithm. These parameters are regarded as part of the fundamental constants of the SM. They are primarily relevant to neutrino physics, but neutrino masses are not considered in this paper since, at present, only upper limits on their values exist. In addition, the experimental precision of the PMNS mixing angles remains relatively low \cite{Esteban:2024eli}, with uncertainties ranging from 5\% to 30\%, which is significantly worse than for the other SM constants.
Including such poorly determined quantities would generate a large number of ``noisy'' analytic solutions in the GP analysis, making them difficult to evaluate within the limited CPU resources available for this study, and potentially obscuring more meaningful relations. Nevertheless, the PMNS mixing angles may be incorporated in future work when greater computational resources become available.

We also did not include cosmological constants and the speed of light ($c$).  However, one should mention that $c$ is indirectly included in the calculation via the fine-structure constant  $\alpha$.

\begin{table}[ht]
  \begin{center}
    \caption{
 Physics fundamental constants from the PDG  \cite{ParticleDataGroup:2024cfk} used for the GP algorithm. 
     We require that the uncertainties for $m_e$, $\pi$ and $\alpha^{-1}$ are at most  a factor 100 higher than for the Higgs boson mass. 
     Making them more precise should not contribute to the results. 
     The second half of the table shows the masses of the fundamental particles in MeV units.  
    The table presents both the absolute uncertainties ($\pm \varepsilon$) and the relative uncertainties $(\pm \varepsilon^{rel}$, in percent). 
    When the PDG reports an asymmetric uncertainty, the largest value is used.
    }
    \label{tab:table1}
    \renewcommand{\tabcolsep}{0.5cm}
    \begin{tabular}{l|r|r|r|r} 
      \hline
      \textbf{Constant} & \textbf{Name} & \textbf{Value} & \boldmath{$\pm \varepsilon$} & \boldmath{$\pm \varepsilon^{rel}$} \textbf{(\%)}\\
      \hline
PI & $\pi$ & 3.14159 & $10^{-5}$ & 0.0003\\
Fine-struct. (inv) & $\alpha^{-1}$ & 137.036 & 0.001 & 0.0007\\
$\alpha_s$ at $Z^0$ & $\alpha_S$ & 0.1180 & 0.0009 & 0.7627\\
\hline
  \textbf{CKM constants}   &    &  \textbf{no units}  & \boldmath{$\pm \varepsilon$} &\boldmath{ $\pm \varepsilon^{rel}$} \textbf{(\%)}\\
\hline
12-mix angle & $\theta_{12}$ & 0.22501 & 0.00068 & 0.3022\\
23-mix angle & $\theta_{23}$ & 0.04183 & 0.00079 & 1.8886\\
13-mix angle & $\theta_{13}$ & 0.003732 & $9\times 10^{-5}$ & 2.4116\\
CP-viol. phase & $\delta$ & 1.147 & 0.026 & 2.2668\\
\hline
  \textbf{Particle masses}   &   & \textbf{MeV} & \boldmath{$\pm \varepsilon$} & \boldmath{$\pm \varepsilon^{rel}$} \textbf{(\%)}\\
\hline
electron mass & $m_e$ & 0.510998 & $10^{-6}$ & 0.0002\\
muon mass & $m_{\mu}$ & 105.658 & 0.001 & 0.0009\\
$\tau$ mass & $m_{\tau}$ & 1776.93 & 0.09 & 0.0051\\
$u$-quark mass & $m_u$ & 2.16 & 0.07 & 3.2407\\
$d$-quark mass & $m_d$ & 4.70 & 0.07 & 1.4894\\
$s$-quark mass & $m_s$ & 93.5 & 0.8 & 0.8556\\
$c$-quark mass & $m_c$ & 1273.0 & 4.6 & 0.3614\\
$b$-quark mass & $m_b$ & 4183 & 7 & 0.1673\\
$t$-quark mass & $m_t$ & 172,560 & 310 & 0.1796\\
$Z$-boson mass & $m_Z$ & 91,188.0 & 2.0 & 0.0022\\
$W$-boson mass & $m_W$ & 80,369.2 & 13.3 & 0.0165\\
$H$-boson mass & $m_H$ & 125,200 & 110 & 0.0879\\
\hline
    \end{tabular}
  \end{center}
\end{table}

The input to the GP algorithm includes two dimensionless constants whose precision significantly surpasses that of the SM particle masses: the mathematical constant $\pi$ and the fine-structure constant $\alpha$ (or its inverse value  
$\alpha^{-1}$, which will be used in our study). Although the constant $\pi$ is not a parameter of the SM, it frequently appears in the laws of physics.
To maintain consistency, we reduce their precision to $\pi=3.14159(1)$ and 
$\alpha^{-1}=137.036(1)$. 
Increasing the precision of $\pi$ and $\alpha^{-1}$ is not expected to affect the results involving particle masses, as the overall uncertainty will still be dominated by uncertainties in the mass values. However, using overly precise constants can introduce numerical instability and make it more difficult for GP to reproduce known relationships. These validation tests will be discussed later.

The main limitation of the inputs that contain a mix of constants with and without physical units is that only relations that pass dimensional analysis have physical meaning. There is no native support for dimensional analyses in the GP. Thus, the created GP output needs to be filtered out to accept only relations where physical units are the same on the left- and the right-hand sides of the relations. This results in significant inefficiency in the current algorithm's generation of useful analytic relations.

The second challenge for this algorithm lies in the wide range of SM values, spanning several orders of magnitude---from the electron mass to the top quark mass, the latter being the heaviest among all SM particles. 
The uncertainties associated with the SM parameters vary significantly and must be taken into account when evaluating the robustness of any discovered relationships. If one variable has very small uncertainties, it offers little advantage if other input variables have large uncertainties, which will dominate the overall uncertainty on the output values. The least precise masses among our inputs are those of the light-flavor quarks ($u$, $d$, $s$), which were taken from the PDG using the $\overline{\text{MS}}$ scheme at a renormalization scale of 2 GeV.
The most significant challenge is also when using the CKM mixing angles as inputs. Due to their relatively large uncertainties, there is a considerable risk of generating spurious equations.

It is important to note that high numerical precision does not necessarily translate to theoretical significance. For example, although the number $\pi$ can be calculated at an extraordinary number of decimal places, it is still just an irrational number, and the usage of all its digits is not essential for a physical theory, since digits beyond a certain level of precision have no observable effect on physical processes.

\subsection{Inputs Without Physical Units}

For the second set of GP inputs, we converted all masses shown in Table~\ref{tab:table1} to values without physical units.
There are arguments \cite{Duff_2014} suggesting that the laws of physics should be independent of the choice of units or measuring instruments. Dimensionless constants, in particular, are well-suited for probing the deep structure underlying physical laws. 

One way to address the physical units issues is to normalize all particle masses by dividing them by another parameter that has a unit of mass. However, this approach introduces a degree of arbitrariness in the choice of the reference mass. A possible option for the divider is the Planck mass  \cite{ParticleDataGroup:2024cfk},  $m_{\mathcal P} = 1.22089(6)\times 10^{19}$~GeV.  While theoretically appealing, this choice is impractical for our purposes, as the GP algorithm we employ cannot handle values smaller than $10^{-17}$.

To resolve this, we normalize all SM masses by the mass of the $\rho (770)$ meson, $775.26 \pm 0.23$ MeV \cite{ParticleDataGroup:2024cfk}. 
At first glance, this might seem like an unconventional choice. One could also use the mass of any fundamental particle for normalization, such as the electron mass. However, that particle would then be excluded from our GP analysis. Our choice is not without justification for the GP algorithm.
The $\rho(770)$ meson holds a special place in the context of quantum chromodynamics,  plays a key role in hadronic structure, and even electromagnetic interactions. Its mass lies conveniently in the GeV range, which allows us to rescale the SM mass spectrum to a range that is well-suited for GP.

Finally, using the uncertainty of the $\rho(770)$ meson in error propagation, when dividing the particle masses by its mass, results in a consistent normalization of uncertainties across all mass values. For GP, it is advantageous to work with uncertainties that are approximately uniform across all inputs, as this improves the stability and comparability of the algorithm’s performance.

For example, the mass of the electron, 0.51099895000(15)~MeV, has an exceptionally small relative uncertainty of $3\times 10^{-8}\%$, while the Higgs boson mass,  $125.20(11)$~GeV \cite{ParticleDataGroup:2024cfk}, has a much larger relative uncertainty of approximately 0.09\%. This disparity in precision poses a challenge for numerical algorithms. 
However, scaling all masses by the $\rho(770)$ meson mass and propagating its uncertainty to the resulting ratios helps mitigate this problem. After this division, the values used for $\pi$ and $\alpha^{-1}$ remain 40 to 100 times more precise than the rescaled electron mass by the $\rho$-meson  mass.

Table~\ref{tab:table1N} presents the masses of the SM particles normalized by the mass of the $\rho$-meson. For consistency, we use the same symbol $m$ to denote the rescaled masses as for the original masses (with units). Thus, for the second set of inputs, we use the constants from Table~\ref{tab:table1}  but where all masses are replaced by the values listed in Table~\ref{tab:table1N}.

\begin{table}[ht]
  \begin{center}
  \caption{
   The masses  \cite{ParticleDataGroup:2024cfk} of the fundamental particles are rescaled to obtain dimensionless values, eliminating the mass units (MeV).
   For this purpose, all particle masses are divided by the mass $775.26 \pm 0.23$ MeV of the $\rho$ meson to improve the value range and reduce the large variability of uncertainties,  making them more uniform for GP. The traditional symbol $m$ was chosen for the rescaled masses to simplify the presentation of the results. The table presents both the absolute measured uncertainties ($\pm \varepsilon$) and the relative uncertainties $(\pm \varepsilon^{rel}$, in percent).  }
    \label{tab:table1N}
     \renewcommand{\tabcolsep}{0.55cm}
    \begin{tabular}{l|r|r|r|r} 
    \hline
      \textbf{Constant} & \textbf{Name} & \textbf{Value} & \boldmath{$\pm \varepsilon$} & \boldmath{$\pm \varepsilon^{rel}$} \textbf{(\%)}\\
      \hline
electron mass & $m_e$ & 0.0006591 & 2 $\times\,10^{-7}$ & 0.0303\\
muon mass & $m_{\mu}$ & 0.1363 & 4 $\times\,10^{-5}$ & 0.0293\\
$\tau$ mass & $m_{\tau}$ & 2.292 & 0.0007 & 0.0305\\
$u$-quark mass & $m_u$ & 0.002786 & 9 $\times\,10^{-5}$ & 3.2304\\
$d$-quark mass & $m_d$ & 0.006062 & 9 $\times\,10^{-5}$ & 1.4847\\
$s$-quark mass & $m_s$ & 0.1206 & 0.0010 & 0.8292\\
$c$-quark mass & $m_c$ & 1.642 & 0.006 & 0.3593\\
$b$-quark mass & $m_b$ & 5.3956 & 0.0092 & 0.1705\\
$t$-quark mass & $m_t$ & 222.583 & 0.405 & 0.1820\\
$Z$-boson mass & $m_Z$ & 117.622 & 0.035 & 0.0298\\
$W$-boson mass & $m_W$ & 103.667 & 0.035 & 0.0338\\
$H$-boson mass & $m_H$ & 161.494 & 0.149 & 0.0923\\
\hline
    \end{tabular}
  \end{center}
\end{table}

We expect the number of valid relations reported by GP for inputs without units to be significantly larger than for inputs with physical units. This is primarily because the output and input variables are unitless. 
Therefore, all relations identified by GP will be considered valid.
In addition, the inputs with the rescaled masses have significantly more uniform relative uncertainties.

It should be emphasized that the presence of the mass of $\rho(770)$ (or any other particle mass used for obtaining dimensionless SM constants) should be considered a handy method for deriving short relationships in GP. It is an auxiliary parameter. The final theory can be free of this constant after successful variable substitutions, which should be performed after obtaining the analytic expressions.
As we discuss later, this means that the dimensionless input allows for the generation of analytically more complex and, at the same time,  dimensionally consistent  expressions. This will lead to a large functional coverage of the potential relations with relatively low CPU usage during GP production.

\subsection{The GP Search Algorithm}

The input parameters discussed above will be used for the following studies as described below. 

First, this paper aims to identify the simplest analytic relationships of the form $f(a)=b$ and $F(a,b)=c$. By ``simplest'', 
we refer to functions $f$ and $F$ with minimal structural complexity, 
or, in most cases, the shortest possible expression length.
We assume that the laws of nature tend to manifest through the simplest possible analytic relationships.

When searching for analytic dependence between constants, one may encounter very complex solutions. To manage this complexity, we have organized all functions according to their simplicity ranks. The lower the rank, the simpler the solution.  To construct the rank for each analytic expression, each mathematical operator was assigned a value as shown in Table~\ref{tab:table2}. These values were summed up to create a rank for the final analytic expression.   For example, $\delta = \pi - 2$ has a rank of 6 because it includes the constant $\pi$ (contributing 1 to the rank), a plain integer (value 2) and a subtraction operation (contributing 3 to the rank).

\begin{table}[ht]
\begin{center}
    \caption{ Assigned values for evaluating ranks for the obtained analytic expressions. These values will be sum up for evaluating the rank of each analytic solution. }
    \label{tab:table2}
\begin{tabular}{lllll}
\hline
 \textbf{Value} & \textbf{Mathematical Token}   \\ \hline
 1 &   physics constant (float) and $\pi$\ \\
 2 &  plain integer ($1\ldots10$) \\
 3 &  addition and subtraction \\
 4 &   multiplication and division \\
 5 &   $\sqrt{x}$,  $x^2$, $x^3$ and $x^4$ \\
 6 &  $\exp(x)$ and $\log(x)$  ($\log$ base of $e$ of $x$), $\cos(x)$, $\sin(x)$,
 $\tan(x)$, $\arctan(x)$ \\ \hline
\end{tabular}
\end{center}
\end{table}

Under this agreement for defining analytic ranks, the algorithm prioritizes basic algebraic operations (such as addition, subtraction, multiplication, and division) while assigning larger values to exponential, logarithmic, and trigonometric functions. Note that the analytic ranks depend on the default representations in the {\sc Picat} output. No attempt is made to simplify expressions.

We will refrain from making additional theoretical assumptions, aiming to keep the solutions as open to potential theories as possible. Even if no apparent structure emerges that links most of the SM parameters, these snippets of relationships may still prove valuable for theory builders, when they are combined together.  They could serve as building blocks for discovering more intricate laws that potentially unify some or all of these fundamental constants.

\subsection{Validation of the GP Algorithm}

As a sanity check, we have validated the genetic programming approach using known relationships. Although these relationships are not discussed in the context of the SM or any theory, they were  established in the past, and have been deemed well suited for our purpose of testing the algorithm.

The first relationship is \cite{1.3022455}:
\begin{equation}
    4 \pi^3 + \pi^2 + \pi  = \alpha^{-1}.
    \label{rq1}
\end{equation}
This expression is particularly useful since both parameters are the most precise in our inputs, thus it provides a reliable test of the GP algorithm. This relation should be true within the precision chosen by us in Table~\ref{tab:table1}. 

The second benchmark test uses the relation between the $Z$ boson mass ($m_Z$), 
the top-quark mass ($m_t$) and the Higgs boson mass ($m_H$) \cite{Torrente-Lujan:2015jea}:

\begin{equation}
    \sqrt{m_Z \>  m_t} = m_H.
    \label{rq2}
\end{equation}

There is no immediate explanation for why such relationships hold, which may suggest that they are merely coincidental.
However, if our calculations are able to reproduce them, we will consider the program to be sufficiently robust.

Our program detected Equations~\eqref{rq1} and \eqref{rq2} expressions using special test runs.  The obtained simplicity ranks were 30 and 20, respectively. The difference $m_H -\sqrt{m_Z \>  m_t}$ = $-0.310$ is greater than the absolute uncertainty allowed on the Higgs mass. When GP finds this equation, it is not considered a solution. 
\section{Results of the GP Algorithm}
\label{sec:results}

In this section, we discuss the output of the GP algorithm.
In total, more than 1 million relations were generated up to simplicity rank 50 for each input, both with and without physical units.
All of these relations are available at \cite{GitRepo}.

\subsection{Output for Inputs with Physical Units}

First, GP was used with the inputs from Table~\ref{tab:table1}. 
Since the input includes a mix of variables with and without physical units, the vast majority
of the resulting relations lack physical meaning.
The dataset containing the GP outputs, along with markers indicating whether each expression satisfies dimensional analysis, is available in \cite{GitRepo}.
These markers were added by an external post-processing program that is not part of the GP framework.
By applying this dimensional-analysis algorithm and visually inspecting expressions with a low analytic rank, we find that approximately 12\% of the generated relations are dimensionally consistent.

Several of the simplest expressions that pass dimensional analysis will be discussed later.

\subsection{Output for Dimensionless Inputs}

We also generated GP snippets using Table~\ref{tab:table1}, after replacing all masses with the dimensionless variables listed in Table~\ref{tab:table1N}.
All expressions in the resulting GP output are dimensionally consistent, offering significant statistics and flexibility for future analyses without requiring additional dimensional analysis. Therefore, in this section, we provide several tables with such analytic snippets that demonstrate the GP output.

\begin{center}
\begin{longtable}{|l|l|r|}
\caption{Analytic expressions for the rank 6. The values of $\Delta D$ show the relative difference between the predicted and target values (in percent).} 
\label{tab:long6} \\
\hline \multicolumn{1}{|c|}{\textbf{Nr}} & \multicolumn{1}{c|}{\textbf{Expression for rank 6}} & \multicolumn{1}{c|}{$\Delta D$ (\%)} \\ \hline 
\endfirsthead
\multicolumn{3}{c}%
{{\bfseries \tablename\ \thetable{} -- continued from previous page}} \\
\hline \multicolumn{1}{|c|}{\textbf{Nr}} & \multicolumn{1}{c|}{\textbf{Expression for rank 6}} & \multicolumn{1}{c|}{$\Delta D$ (\%)} \\ \hline 
\endhead
\hline \multicolumn{3}{|r|}{{Continued on next page}} \\ \hline
\endfoot

\hline \hline
\endlastfoot
1 & $\alpha_S = m_{\mu} / \delta$ & 0.71 \\ 
2 & $\delta = 1 + m_{\mu}$ & 0.93 \\ 
3 & $\delta = \pi - 2$ & 0.47 \\ 
4 & $\delta = m_{\mu} / \alpha_S$ & 0.70 \\ 
5 & $\theta_{23} = \theta_{12} / m_b$ & 0.30 \\ 
\end{longtable} 
 \end{center}
\begin{center}
\begin{longtable}{|l|l|r|}
\caption{Analytic expressions for the rank 7. The values of $\Delta D$ show the relative difference between the predicted and target values (in percent).} 
\label{tab:long7} \\
\hline \multicolumn{1}{|c|}{\textbf{Nr}} & \multicolumn{1}{c|}{\textbf{Expression for rank 7}} & \multicolumn{1}{c|}{$\Delta D$ (\%)} \\ \hline 
\endfirsthead
\multicolumn{3}{c}%
{{\bfseries \tablename\ \thetable{} -- continued from previous page}} \\
\hline \multicolumn{1}{|c|}{\textbf{Nr}} & \multicolumn{1}{c|}{\textbf{Expression for rank 7}} & \multicolumn{1}{c|}{$\Delta D$ (\%)} \\ \hline 
\endhead
\hline \multicolumn{3}{|r|}{{Continued on next page}} \\ \hline
\endfoot

\hline \hline
\endlastfoot
1 & $\delta = \exp(m_{\mu})$ & 0.08 \\ 
2 & $\delta = \log(\pi)$ & 0.20 \\ 
3 & $\delta = m_{\tau} / 2$ & 0.09 \\ 
\end{longtable} 
 \end{center}

As an example, the expressions with the lowest analytic ranks, 6 and 7, are shown in Tables~\ref{tab:long6} and \ref{tab:long7}, respectively.
The regularities appear rather trivial, and can be checked by hand. Table~\ref{tab:long10} of the Appendix~\ref{appendix} shows the results with rank 10. 
Note that the analytic expressions with ranks 8 and 9 do not exist because of the default construction of the ranking system.
We intentionally retained repeated snippets involving the same variables, as they may help model builders identify interesting patterns.

The tables also show the relative difference for each analytic snippet.
It is defined as $\Delta D= 100\% \times |y-y_{t}|/y_t$ (expressed as a percentage),
where $y_t$ is the target value and $y$ is the predicted value. All relations have
$\Delta D<\varepsilon^{rel}$ on the target constants according to the definition of GP.  The distribution of the $\Delta D$ values will be discussed later.
Only formulas with $\Delta D$ smaller than 1\%  are listed 
 to reduce the sizes of the tables. This requirement truncates the presentation for some targets, which usually correspond to $u,d$-quark masses, the CKM mixing angles and CP-violating phase, since they have relative uncertainties $\varepsilon^{rel}$ above 1\%; see Table~\ref{tab:table1N}.  Other target constants should not be affected by the 1\% requirement, since their $\Delta D$ values are smaller than $\varepsilon^{rel}$ listed in Table~\ref{tab:table1}.

The results shown in Tables~\ref{tab:long6} and \ref{tab:long10} are dominated by the expressions for $\delta$, which has the largest absolute and relative uncertainties ($\varepsilon$ and $\varepsilon^{rel}$) among all the input variables.
Among all the ranks obtained, the lepton masses had the fewest solutions due to their relatively small mass uncertainty. All relationships from the symbolic regression have been preserved, even if they may appear highly unlikely. We remind that this choice reflects our intent to remain as theory-agnostic as possible, in case these analytic snippets are combined and simplified, and later used to inform or construct a model that connects all SM constants.

Most of the obtained GP solutions, if not all, are likely coincidental and represent numerical noise. Identifying meaningful dynamic relationships based on symmetry or other mathematical and physical principles remains an important direction for future work. A visual inspection of the lowest-rank expressions revealed no notable patterns, such as systematically repeated constants or functions, that might indicate underlying regularities. The frequencies of integers from 1 to 10 appear to be uniformly distributed, with approximately the same mean value. 

We should recall that the introduction of the non-fundamental particle, the  $\rho(770)$ meson, into our study was made to provide dimensionless inputs for GP, which are essential for the computational efficiency. The mass of this particle is an auxiliary parameter. It defines the limits of numerical precision for the GP and, consequently, the applicability of any potential theory connecting the SM constants, should such a theory exist.
In principle, any particle can be used for this purpose, as long as the rescaled masses remain within a numerically suitable range for the GP framework. The analytic relations using the 
$\phi(1020)$-meson mass, $1019.461 \pm 0.016$~MeV \cite{ParticleDataGroup:2024cfk}, are available from \cite{GitRepo}.

\begin{figure}[th]  
  \begin{center}
     \includegraphics[width=0.8\textwidth]{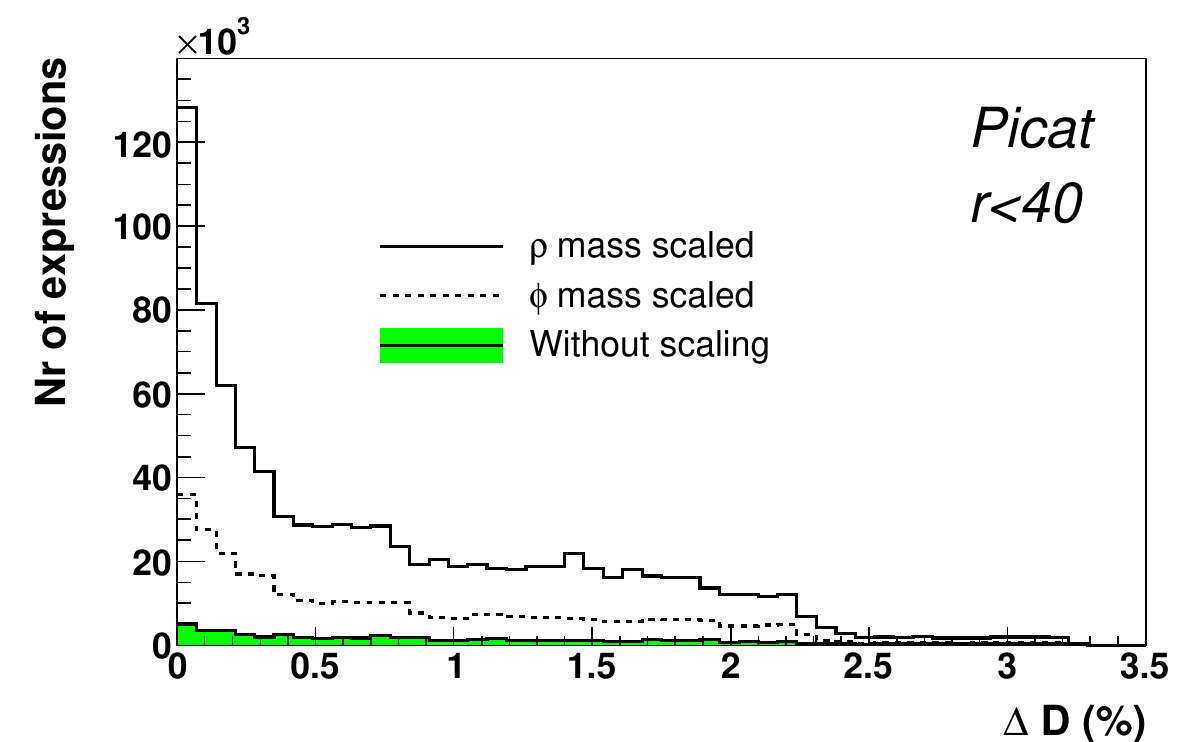}
  \end{center}
\caption{The distribution of $\Delta D= 100\% \times |y-y_{t}|/y_t$ (see the text) for the expressions without mass scaling (the filled histogram), and after scaling by the $\rho$ and
$\phi$ meson masses. Only expressions with analytic rank smaller than 40 are shown.}
\label{plot1}
\end{figure}

As a measure of how many expressions can be used for further study, Figure~\ref{plot1} shows the distribution of the values of $\Delta D$ (as defined above) for the GP outputs for the SM constants without scaling (filled histogram) and with the scaling by the  $\rho$ and $\phi$ masses. Only expressions with an analytic rank of less than 40 are shown. A cutoff of 40 was chosen to avoid overly complex expressions. 

As one can see from  Figure~\ref{plot1}, the number of GP expressions with scaled SM constants is significantly larger than the number of relations without mass scaling. 
We also recall that only 12\% of the expressions without mass scaling pass dimensional analysis. A smaller number of GP expressions for the $\phi$-mass normalization, compared to the $\rho$-mass normalization, is due to the 
fact that the $\phi$-meson mass has a smaller relative experimental uncertainty 
than the  $\rho$-meson mass.
\vspace{-6pt}

Note that all rescaled masses in higher-rank analytical expressions can be straightforwardly converted to actual masses in units of GeV or MeV using the most suitable relation from Table~\ref{tab:long6} or Table \ref{tab:long7}, 
followed by appropriate substitutions.
For example, one can use $\delta = \exp (m'_{\mu} / m'_{\rho})$ from Table~\ref{tab:long7}, where $m'_{\mu}$ and $m'_{\rho}$ denote the muon and $\rho(770)$ meson masses in MeV. From this expression, the value of $m'_{\rho}$ can be determined and subsequently applied across all dimensionless expressions derived in this study. This means that the dimensionless input allows for the generation of high-ranking, dimensionally consistent analytical expressions with relatively low CPU usage during GP production.
Determining the optimal analytic expression for the $\rho(770)$ mass from the GP outputs, however, 
lies beyond the scope of this paper. 
This task should be pursued by model builders in future analyses of the derived analytic snippets, with the goal of identifying relationships among fundamental constants that enable the construction of physics models with a minimal set of parameters.

\subsection{Discussion of the Results}

In this section, we discuss the outputs generated by the GP algorithm that are common to both dimensional and dimensionless inputs. Outputs that satisfy dimensional analysis are more easily obtained when using dimensionless inputs, achieved by rescaling the masses. We recall that rescaling reduces the disparity in numerical values, particularly when considering lepton masses. The discussion below serves as an example of the step (2) described in Section~\ref{sec:method}. A complete set of requirements for this stage of the analysis should be addressed in a dedicated physics paper, which is beyond the scope of this work.

A preliminary visual inspection of the lowest‑rank expressions reveals relationships that appear to be both remarkably simple and precise, warranting further investigation.
For example,  Table~\ref{tab:long10} has an expression $m_t = m_W + m_W \> \delta$ 
connecting the top-quark mass $m_t$ with the $W$ mass and $\delta$. 
Tables with analytic ranks above 10 are available from \cite{GitRepo}.
The rank 15 snippets listed in \cite{GitRepo} contain another relation
$m_H = m_t - (m_b + m_t / 4)$
connecting the Higgs mass $m_H$ with $m_t$ and the $b-$quark mass. Another relation of the same rank is $m_u = 2 m_e/\sqrt{\theta_{12}}$, where $m_u$ is the mass
of the $u$-quark, $m_e$ is the mass of the electron and $\theta_{12}$ is the CKM 12 mixing angle.
The found expression $m_H = 5\> m_W  - 3 \> (m_t - m_W)$  connects 
the Higgs mass with the top-quark mass and the $W$ mass. 
All such expressions are valid for the original masses with  units, without scaling by the $\rho(770)$ mass.
The precision of these analytic relations significantly exceeds that of Equation~\eqref{rq2}.

We should mention that many relations involving the Higgs mass are valid only for the dimensionless constants listed in Table~\ref{tab:table1N}. For example, $m_H = m_t - 6\> \sqrt{m_W}$ connects $m_H$, $m_W$ and $m_t$. It does not pass the dimensional analysis when using the original masses, i.e., without rescaling the masses by the $\rho(770)$ meson mass. However, the mass of this meson can easily be substituted using Table~\ref{tab:long6} or Table \ref{tab:long7}.

To identify meaningful relationships from the GP output, it is necessary to apply constraints that reduce the number of relations arising from mere numerical coincidences. For the SM, such a constraint could be defined as follows:
One of the known properties of the SM is that the masses of fermions, as well as the $m_Z$ and $m_W$ boson masses, tend toward zero when the Higgs boson mass is set to zero. This behavior arises from their dependence on the Higgs vacuum expectation value through the Yukawa couplings, which are free parameters in the theory \cite{PhysRevLett.19.1264}. We identified several simplest relations connecting 
the quark and $Z/W$ masses with the Higgs mass, which have the above-mentioned property, and  also pass the dimensional analysis. Here are some examples with small analytic ranks:

\begin{align}
m_u &=\frac{m_H}{3\> (\alpha^{-1})^2},   \quad m_d = m_u  (1 + \delta), \\
m_s &=  \frac{m_d} {\theta_{12}^2},   \quad  m_c =  m_d\>  \left(2+ \frac{1}{\theta_{13}}\right),  \\	
m_b &=  \frac{m_c \> \pi^2}{3},  \quad  m_t = m_b \> \pi (\pi +10), \\
m_Z &= m_t \> \log(\delta)  (5 - \delta), \quad m_W = m_t / (1+\delta).
\label{eq2}
\end{align}
These relations hold for the masses presented in Table~\ref{tab:table1} and for the rescaled masses in Table~\ref{tab:table1N}.
Analytical expressions for lepton masses as linear functions of the Higgs mass were not found. Nevertheless, these examples illustrate that the reduction of the free parameters of the SM is possible, since 8 masses are expressed via the Higgs mass and 4 measured constants ($\alpha^{-1}$, $\delta$, $\theta_{12}$ and $\theta_{13}$).  The origin of these expressions may be purely coincidental, and can only be understood within the framework of a specific model, which is beyond the scope of this paper. 

The obtained analytic expressions can be used to develop theories by model builders, or serve as inputs for various artificial intelligence techniques to discover hidden analytic structures. For example, the analytic snippets can be simplified or combined, variables can be substituted, and dimensional analysis can be applied after converting the masses back to their original physical units.

\section{Next Steps}

The goal of this paper is to explore the first step in the physics program: identifying the analytic structures among SM parameters. This objective is outlined in Section~\ref{sec:method}. Two approaches are discussed in this paper: (1) Using the original SM parameters with mixed units, followed by the application of post-processing algorithms to verify the dimensional correctness of the obtained expressions;
(2) Generating dimensionally consistent expressions after normalizing by an auxiliary mass.
The first approach produces smaller numerical noise (after dimensional analysis), while the second approach results in a significant number of dimensionally correct expressions, offering greater flexibility for model builders. However, in the latter case, the challenge shifts from the GP itself to the analytic substitutions necessary to remove the dependence on the auxiliary mass. It should be noted that there is significant overlap between these two approaches, since dimensionally correct expressions from method (1) can also be found among the expressions of method (2). 
We expect that the numerical noise is higher in the second approach.

On the technical side, several improvements can be made for GP, such as incorporating variations of arguments in the discovered expressions and performing dimensional checks inside the GP algorithm. The latter may improve the efficiency of GP, since external algorithms for checking dimensional correctness would no longer be required. 

The inclusion of cosmological constants is another possible extension, which can be implemented in a straightforward manner, since the analysis code is publicly available. However, one should keep in mind the limitation of the {\sc Picat} program: it cannot handle values smaller than $10^{-17}$. For this reason, this paper did not use the Planck mass, which is more natural in particle physics for removing mass units.

Even when dimensionally consistent expressions are identified, this does not completely eliminate the numerical noise arising from coincidental relations. The second step, as discussed in Section~\ref{sec:method}, must therefore involve filtering out coincidental expressions with the goal of identifying relations that can be given a physical interpretation within some unification model. Although this step is beyond the scope of the present paper, here, we should mention one statistical method that can be used to address this problem.  
Recently, an approach called the ``random sampling test'' was introduced~\cite{Chekanov:2025pmb}, in which random smearing is applied to the GP inputs followed by re-evaluation of the simplicity ranks. 
If a consistent system of the relations exhibits the lowest simplicity rank among all random variations, it may suggest a genuine interdependence among the fundamental constants within a high-dimensional functional space.
This method was found to be CPU-intensive, since every variation of the GP inputs  requires the creation of new analytic snippets, followed by dimensional analysis, and re-calculation of simplicity ranks. To address high CPU usage in the future, large-scale high-performance computing resources may be required.

It is important to emphasize that any meaningful physical model or theory must be falsifiable.
This means that as measurements of SM constants become more precise, the GP relations used by such a model or theory can be retested; if they fail to hold, that mathematical framework is not viable. In this sense, the numerical noise problem is itself falsifiable.

\section{Conclusions}
\label{sec:conclusion}

This paper discusses a promising method based on genetic programming for uncovering underlying mathematical relationships directly from data. The first step follows a  symbolic regression approach: deriving analytical expressions from the data and generating a large set of candidate relations. The second step involves identifying simplest connecting patterns among these relations, dominated by numerical coincidences. In this paper, this step leverages dimensional analysis and general SM expectations to reveal possible analytic structures within a high-dimensional functional space. To generalize the use of this method,  the SM constraints can be substituted with any known conditions pertinent to the expected outcome, allowing for the filtering of expressions that are likely artifacts of numerical noise.

The derived analytic expressions---referred to as ``snippets''---connect the fundamental constants of the SM using an approach based on GP. 
We presented results up to analytic rank 10, limiting the solutions to a relative precision of at most 1\%.  The results with a relative difference greater than 1\%, corresponding to the $u$- and $d$-quark masses, the CKM mixing angles,  and the CP-violating phase, are not included in this article. Analytic expressions with ranks above 10, without any restriction on predicted precision $\Delta D$, and with and without rescaling by the $\rho(770)$ (or $\phi(1020)$) meson mass, are available \cite{GitRepo} from the authors. The same repository includes a basic example with the analysis code and the configuration files for the {\sc Picat} program \cite{PicatWebsite}.
  
A few known (though likely coincidental) relationships were successfully recovered. 
Surprisingly, the obtained GP results reveal a significant number of simple analytic patterns. Some of them can pass dimensional analysis. 
However, no evident patterns emerged from the relationships between the constants that could indicate an underlying simple mathematical structure with a limited number of free parameters. Although we expect that most of the expressions, undoubtedly,  arise by pure chance, it is possible that some may hint at an underlying structure or deeper theory linking these fundamental constants---should such a theory exist. The coincidental nature of some (or all) of the snippets will be revealed over time as more precise data on the SM parameters become available.

We believe that the compiled lookup library of analytic snippets will be useful for model builders.
We also hope that the derived expressions will serve as a valuable resource for future AI-driven investigations, providing building blocks for uncovering a potential underlying law that connects all parameters of the SM through a small set of fundamental constants. 
Finally, the proposed method can be extended to other research fields characterized by numerous free parameters, such as the study of cosmological constants and various branches of applied physics.


\section*{Acknowledgments}
We would like to thank Dr.~C.~K.~Zachos for valuable discussions. 
The submitted manuscript has been created by UChicago Argonne, LLC, Operator of Argonne National Laboratory (“Argonne”). Argonne, a U.S. 
Department of Energy Office of Science laboratory, is operated under Contract No. DE-AC02-06CH11357.  Argonne National Laboratory’s work was 
funded by the U.S. Department of Energy, Office of High Energy Physics under contract DE-AC02-06CH11357.

\newpage

\bibliographystyle{JHEP}
\bibliography{references}

@article{ParticleDataGroup:2024cfk,
    author = "Navas, S. and others",
    collaboration = "{Particle Data Group}",
    title = "{Review of particle physics}",
    doi = "10.1103/PhysRevD.110.030001",
    journal = "Phys. Rev. D",
    volume = "110",
    number = "3",
    pages = "030001",
    year = "2024"
}

@article{Chekanov:2025pmb,
    author = "Chekanov, S. V. and Kjellerstrand, H.",
    title = "{Evidence of Relationships Among Fundamental Constants of the Standard Model}",
    eprint = "2509.07713",
    archivePrefix = "arXiv",
    primaryClass = "hep-ph",
    reportNumber = "HEP-ANL-198667",
    month = "9",
    year = "2025",
    note = "HEP-ANL-198667",
}

@article{Esteban:2024eli,
    author = "Esteban, Ivan and Gonzalez-Garcia, M. C. and Maltoni, Michele and Martinez-Soler, Ivan and Pinheiro, Jo{\~a}o Paulo and Schwetz, Thomas",
    title = "{NuFit-6.0: updated global analysis of three-flavor neutrino oscillations}",
    eprint = "2410.05380",
    archivePrefix = "arXiv",
    primaryClass = "hep-ph",
    reportNumber = "IFT-UAM/CSIC-24-140, YITP-SB-2024-24, IPPP/24/64, IPPP/24/64, IFT-UAM/CSIC-24-140, YITP-SB-2024-24",
    doi = "10.1007/JHEP12(2024)216",
    journal = "JHEP",
    volume = "12",
    pages = "216",
    year = "2024"
}

@book{WBook,
   author={Weinberg, S.},
   publisher={Pantheon Books},
   title={{Dreams of a Final Theory}},
   year={1994}
    }

@article{PhysRevLett.19.1264,
  title = {A Model of Leptons},
  author = {Weinberg, Steven},
  journal = {Phys. Rev. Lett.},
  volume = {19},
  issue = {21},
  pages = {1264--1266},
  numpages = {0},
  year = {1967},
  month = {Nov},
  publisher = {American Physical Society},
  doi = {10.1103/PhysRevLett.19.1264},
  url = {https://link.aps.org/doi/10.1103/PhysRevLett.19.1264}
}

@misc{GitRepo,
  author = {Chekanov, S. V. and Kjellerstrand, H.},
  title = {{GC4PhysicalConstants - Genetic computing for physical constants}},
  year = {GitHub repository, 2025},
  publisher = {GitHub},
  note = {GitHub repository},
  howpublished = {\url{https://github.com/chekanov/GC4PhysicalConstants}},
  URL = {https://github.com/chekanov/GC4PhysicalConstants}
}

@Article{universe10110414,
AUTHOR = {Chekanov, Sergei V.},
TITLE = {{Estimation of the chances to find new phenomena at the LHC in a model-agnostic combinatorial analysis}},
JOURNAL = {Universe},
VOLUME = {10},
YEAR = {2024},
NUMBER = {11},
ARTICLE-NUMBER = {414},
URL = {https://www.mdpi.com/2218-1997/10/11/414},
ISSN = {2218-1997},
DOI = {10.3390/universe10110414}
}

@article{Froggatt__2003,
   title={{Trying to understand the standard model parameters}},
   volume={18},
   ISSN={1477-2892},
   url={http://dx.doi.org/10.1080/0142241032000156559},
   DOI={10.1080/0142241032000156559},
   number={1–4},
   journal={Surveys in High Energy Physics},
   publisher={Informa UK Limited},
   author={Froggatt, C. D. and Nielsen, H. B.},
   year={2003},
   month=jan, pages={55–75} }

@inproceedings{Nielsen:1994ab,
    author = "Nielsen, Holger Bech and Surlykke, Christian and Rugh, Svend Erik",
    title = "{Seeking inspiration from the standard model in order to go beyond it}",
    booktitle = "{4th Hellenic School on Elementary Particle Physics}",
    eprint = "hep-th/9407012",
    archivePrefix = "arXiv",
    reportNumber = "NBI-HE-93-48-REV, NBI-HE-93-48",
    pages = "476--501",
    month = "7",
    year = "1994"
}

@article{Duff_2014,
   title={How fundamental are fundamental constants?},
   volume={56},
   ISSN={1366-5812},
   url={http://dx.doi.org/10.1080/00107514.2014.980093},
   DOI={10.1080/00107514.2014.980093},
   number={1},
   journal={Contemporary Physics},
   publisher={Informa UK Limited},
   author={Duff, M.J.},
   year={2014},
   month=dec, pages={35–47} }

@article{1.3022455,
    author = {Roskies, Ralph and Peres, Asher},
    title = {A new pastime–calculating alpha to one part in a million},
    journal = {Physics Today},
    volume = {24},
    number = {11},
    pages = {9-9},
    year = {1971},
    month = {11},
    issn = {0031-9228},
    doi = {10.1063/1.3022455},
    url = {https://doi.org/10.1063/1.3022455}
}

@article{Torrente-Lujan:2015jea,
    author = "Torrente-Lujan, E.",
    editor = "Bravina, L. and Foka, Y. and Kabana, S.",
    title = "{The Higgs and top mass coincidence problem}",
    doi = "10.1051/epjconf/20149505015",
    journal = "EPJ Web Conf.",
    volume = "95",
    pages = "05015",
    year = "2015"
}

@misc{PicatWebsite,
  author       = {Zhou, Neng-Fa and Kjellerstrand, H{\aa}kan and Fruhman, Jonathan},
  title        = {{\sc Picat} Programming Language},
  howpublished = {\url{https://picat-lang.org}},
  year         = {2013--\the\year},
  note         = {[Online; accessed \today]},
}

@misc{PicatSymbolicRegression,
  author       = {Kjellerstrand, H{\aa}kan},
  title        = {symbolic\_regression.pi},
  howpublished = {\url{https://hakank.org/picat/symbolic\_regression.pi}},
  year         = {2025},
  note         = {[Online; accessed \today]},
}

@book{koza92,
  author = {Koza, J. R.},
  publisher = {MIT Press},
  title = {Genetic Programming},
  username = {dalbem},
  year = 1992
}




\clearpage
\appendix
\part*{Appendix}
\addcontentsline{toc}{part}{Appendix}
\label{appendix}
\label{secA}
Table \ref{tab:long10}  lists the analytic 
snippets for the rank 10.
We retain the mathematical notations used in 
the {\sc Picat} language to facilitate future processing.
They  employ round brackets for functions such as $\sqrt{x}$, while the operators $pow2(x)$, $pow3(x)$, $pow4(x)$ refer a power of 2, 3 and 4, respectively.

\begin{center}
\begin{longtable}{|l|l|r|}
\caption{Analytic expressions for the rank 10. The values of $\Delta D$ show the relative difference between the predicted and target values (in percent).} 
\label{tab:long10} \\
\hline \multicolumn{1}{|c|}{\textbf{Nr}} & \multicolumn{1}{c|}{\textbf{Expression for rank 10}} & \multicolumn{1}{c|}{$\Delta D$ (\%)} \\ \hline 
\endfirsthead
\multicolumn{3}{c}%
{{\bfseries \tablename\ \thetable{} -- continued from previous page}} \\
\hline \multicolumn{1}{|c|}{\textbf{Nr}} & \multicolumn{1}{c|}{\textbf{Expression for rank 10}} & \multicolumn{1}{c|}{$\Delta D$ (\%)} \\ \hline 
\endhead
\hline \multicolumn{3}{|r|}{{Continued on next page}} \\ \hline
\endfoot

\hline \hline
\endlastfoot
1 & $\alpha_S = (m_Z - m_W) / m_Z$ & 0.54 \\ 
2 & $\alpha_S = m_e / (m_u + m_u)$ & 0.24 \\ 
3 & $\alpha_S = m_s - pow2(\theta_{23})$ & 0.72 \\ 
4 & $\alpha_S = m_s - pow3(m_s)$ & 0.72 \\ 
5 & $\alpha_S = m_s - pow3(m_{\mu})$ & 0.06 \\ 
6 & $\alpha_S = m_s - pow4(\theta_{12})$ & 0.03 \\ 
7 & $\alpha_S = m_s / (\delta - m_s)$ & 0.43 \\ 
8 & $\alpha_S = m_{\mu} - pow2(m_{\mu})$ & 0.24 \\ 
9 & $\delta = (\theta_{12} + m_c) / m_c$ & 0.87 \\ 
10 & $\delta = (\theta_{23} + m_d) / \theta_{23}$ & 0.18 \\ 
11 & $\delta = (m_t - m_W) / m_W$ & 0.009 \\ 
12 & $\delta = 1 + (m_{\mu} + \theta_{13})$ & 0.61 \\ 
13 & $\delta = 1 + m_{\mu} + m_e$ & 0.88 \\ 
14 & $\delta = 1 + m_{\mu} - m_e$ & 0.99 \\ 
15 & $\delta = 2 + (m_{\tau} - \pi)$ & 0.30 \\ 
16 & $\delta = \alpha^{-1} / \alpha^{-1} + m_{\mu}$ & 0.93 \\ 
17 & $\delta = \alpha_S + m_s / \alpha_S$ & 0.61 \\ 
18 & $\delta = \pi - (2 + \theta_{13})$ & 0.80 \\ 
19 & $\delta = \pi - (2 + m_e)$ & 0.53 \\ 
20 & $\delta = \pi - m_u - 2$ & 0.71 \\ 
21 & $\delta = \theta_{13} + \pi - 2$ & 0.15 \\ 
22 & $\delta = m_b / m_b + m_{\mu}$ & 0.93 \\ 
23 & $\delta = m_c / m_c + m_{\mu}$ & 0.93 \\ 
24 & $\delta = m_d + (\pi - 2)$ & 0.06 \\ 
25 & $\delta = m_d / m_d + m_{\mu}$ & 0.93 \\ 
26 & $\delta = m_e + \pi - 2$ & 0.41 \\ 
27 & $\delta = m_e / m_e + m_{\mu}$ & 0.93 \\ 
28 & $\delta = m_s + m_s / \alpha_S$ & 0.38 \\ 
29 & $\delta = m_s / (\theta_{12} - m_s)$ & 0.70 \\ 
30 & $\delta = m_t / m_t + m_{\mu}$ & 0.93 \\ 
31 & $\delta = m_u - (2 - \pi)$ & 0.23 \\ 
32 & $\delta = m_u / m_u + m_{\mu}$ & 0.93 \\ 
33 & $\delta = m_{\mu} + 1 + m_u$ & 0.69 \\ 
34 & $\delta = m_{\mu} + \alpha_S / \alpha_S$ & 0.93 \\ 
35 & $\delta = m_{\mu} + \pi / \pi$ & 0.93 \\ 
36 & $\delta = m_{\mu} + \theta_{12} / \theta_{12}$ & 0.93 \\ 
37 & $\delta = m_{\mu} + \theta_{13} / \theta_{13}$ & 0.93 \\ 
38 & $\delta = m_{\mu} + \theta_{23} / \theta_{23}$ & 0.93 \\ 
39 & $\delta = m_{\mu} + m_H / m_H$ & 0.93 \\ 
40 & $\delta = m_{\mu} + m_W / m_W$ & 0.93 \\ 
41 & $\delta = m_{\mu} + m_Z / m_Z$ & 0.93 \\ 
42 & $\delta = m_{\mu} + m_d + 1$ & 0.40 \\ 
43 & $\delta = m_{\mu} + m_s / m_s$ & 0.93 \\ 
44 & $\delta = m_{\mu} + m_{\mu} / m_{\mu}$ & 0.93 \\ 
45 & $\delta = m_{\mu} + m_{\tau} / m_{\tau}$ & 0.93 \\ 
46 & $\delta = m_{\tau} - m_{\mu} - 1$ & 0.76 \\ 
47 & $\delta = \sqrt(m_c) - m_{\mu}$ & 0.16 \\ 
48 & $\theta_{12} = \sqrt(\alpha_S) - \alpha_S$ & 0.22 \\ 
49 & $\theta_{13} = m_d / (m_c - m_d)$ & 0.71 \\ 
50 & $\theta_{23} = m_s * (\theta_{12} + m_s)$ & 0.36 \\ 
51 & $\theta_{23} = m_{\mu} / (\pi + m_{\mu})$ & 0.59 \\ 
52 & $\theta_{23} = m_{\mu} / (m_c + m_c)$ & 0.78 \\ 
53 & $m_H = (\theta_{12} + \theta_{12}) / m_u$ & 0.02 \\ 
54 & $m_W = \delta + m_Z / \delta$ & 0.03 \\ 
55 & $m_W = m_H * m_c - m_H$ & 0.01 \\ 
56 & $m_Z = \delta * (m_W - \delta)$ & 0.03 \\ 
57 & $m_b = 7 + (\theta_{23} - m_c)$ & 0.08 \\ 
58 & $m_b = m_c * (m_c + m_c)$ & 0.06 \\ 
59 & $m_b = m_{\mu} + pow2(m_{\tau})$ & 0.11 \\ 
60 & $m_b = pow4(m_c - \alpha_S)$ & 0.02 \\ 
61 & $m_c = (m_H + m_W) / m_H$ & 0.005 \\ 
62 & $m_c = 2 - (m_{\mu} + \theta_{12})$ & 0.20 \\ 
63 & $m_c = 7 + (\theta_{23} - m_b)$ & 0.26 \\ 
64 & $m_c = \delta + \delta / m_{\tau}$ & 0.33 \\ 
65 & $m_c = \theta_{12} / m_s - \theta_{12}$ & 0.08 \\ 
66 & $m_c = pow2(m_{\mu} + \delta)$ & 0.30 \\ 
67 & $m_c = pow3(\delta) + m_{\mu}$ & 0.20 \\ 
68 & $m_c = \sqrt(\pi) - m_{\mu}$ & 0.36 \\ 
69 & $m_d = \theta_{13} * (m_c - \theta_{13})$ & 0.86 \\ 
70 & $m_d = \theta_{13} / m_c + \theta_{13}$ & 0.94 \\ 
71 & $m_s = \alpha_S * (\delta - \alpha_S)$ & 0.68 \\ 
72 & $m_s = \alpha_S + pow2(\theta_{23})$ & 0.70 \\ 
73 & $m_s = \alpha_S + pow3(m_{\mu})$ & 0.06 \\ 
74 & $m_s = \theta_{12} / (\theta_{12} + m_c)$ & 0.07 \\ 
75 & $m_s = \theta_{23} + \sqrt(m_d)$ & 0.76 \\ 
76 & $m_s = m_{\mu} - \alpha_S * m_{\mu}$ & 0.32 \\ 
77 & $m_s = pow3(\alpha_S) + \alpha_S$ & 0.79 \\ 
78 & $m_s = pow3(m_c - \delta)$ & 0.57 \\ 
79 & $m_s = pow4(\theta_{12}) + \alpha_S$ & 0.03 \\ 
80 & $m_t = 1 + (m_Z + m_W)$ & 0.13 \\ 
81 & $m_t = m_W + \delta * m_W$ & 0.004 \\ 
82 & $m_t = m_c * (\alpha^{-1} - m_c)$ & 0.12 \\ 
83 & $m_u = (\theta_{12} + \theta_{12}) / m_H$ & 0.02 \\ 
84 & $m_u = m_e / (\alpha_S + \alpha_S)$ & 0.24 \\ 
\end{longtable} 
 \end{center}



\end{document}